\documentclass[a4paper]{article}
\usepackage{geometry}
\usepackage{amsmath}
\usepackage{amsfonts}
\usepackage{amssymb}
\usepackage{color}
\usepackage[pdftex]{graphicx}
\usepackage{authblk}
\allowdisplaybreaks[4]
\everymath{\displaystyle}

\newcommand\keywords{}

\begin{document}
\title{Asymptotic behavior of a stochastic particle system \\ of 5 neighbors}
\author{Kazushige Endo}
\affil{National Institute of Technology Gunma College, 580, Toribamachi, Maebashi City, Gunma, 371-8530, Japan}
\date{}
\maketitle
\begin{abstract}
We analyze a stochastic particle system of 5 neighbors. Considering eigenvalue problem of transition matrix, we propose a conjecture that asymptotic distribution of the system is determined by the number of specific local patterns in the asymptotic solution. Based on the conjecture, mean flux which depends of a pair of the conserved quantities is derived theoretically. Moreover, we obtain mean flux in the deterministic case through the limit of stochastic parameter.
\end{abstract}

\section{Introduction} \label{1}
Discrete particle systems have been studied for various fields such as physics, engineering, and mathematics. Especially, dependency of the momentum of particles on their density in the asymptotic behavior have been the one of important topics for analyzing particle systems\cite{fuks, nishinari}. For example, asymmetric simple exclusion process (ASEP) is a popular discrete stochastic particle system which is a multibody random walk model where each particle moves stochastically along one-dimensional lattice space\cite{spitzer, derrida1, derrida2}. Its steady state is discussed relating to orthogonal polynomials\cite{sasamoto}. Nagel-Schreckenberg model is another interesting example related to traffic flow\cite{nagel, ns}. Nagel and Schreckenberg theoretically analyzed mechanism of traffic jam, which depends on occupancy of cars in a traffic way.\par
In this article, we investigate the asymptotic dynamics of the stochastic system and discuss the dependency of mean flux on other quantities. Mean flux of ASEP is determined by particle density uniquely. However, that of our system depends not only particle density but also another conserved quantity.\par
\begin{figure}[h]
\begin{center}
\setlength\unitlength{1truecm}
\begin{picture}(4,4.5)(1.5,0)
\put(-0.27,4.3){\vector(1,0){2}}
\put(2,4.2){$j$}
\put(-0.27,4.3){\vector(0,-1){2}}
\put(-0.35,1.93){$n$}
\put(0,0){\includegraphics[width=70mm]{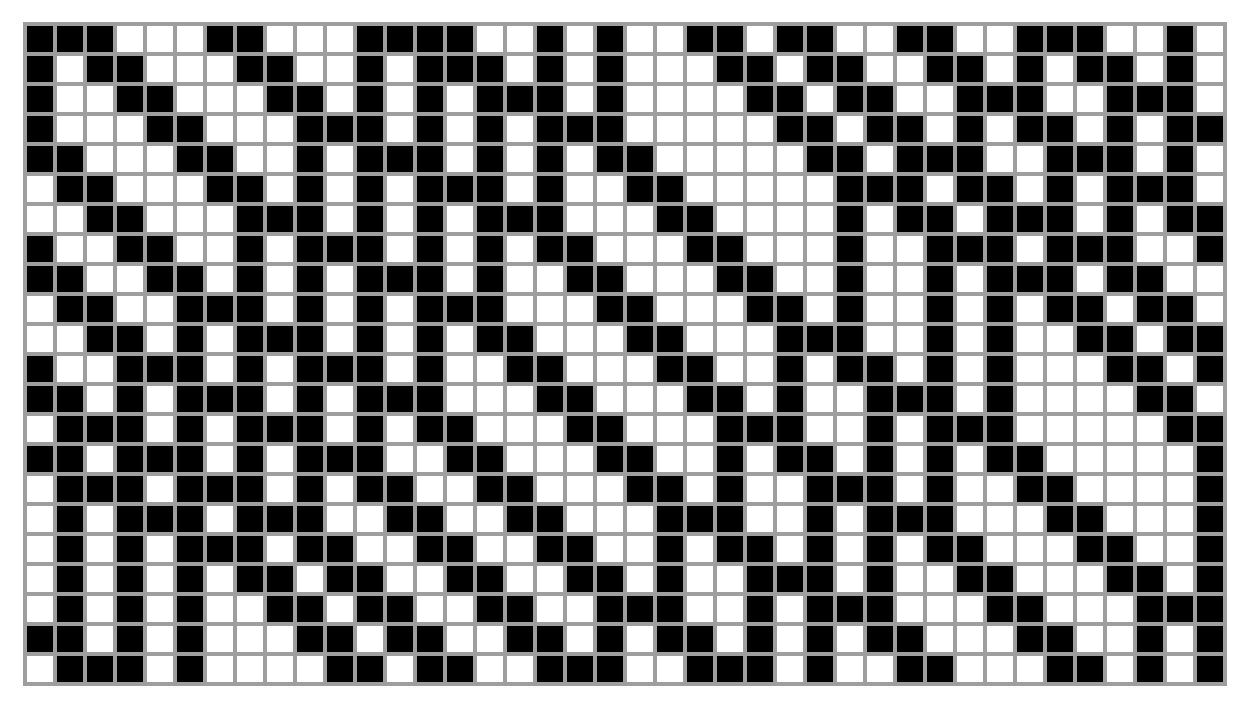}}
\end{picture}
\caption{Example of time evolution of (\ref{dsystem}). Black squares $\blacksquare$ mean $u=1$ and white squares $\square$ $u=0$.}
\label{te1}
\end{center}
\end{figure}
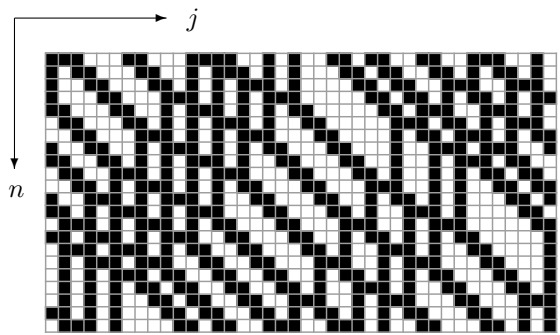
\begin{figure}[h]
\begin{center}
\includegraphics[width=80mm]{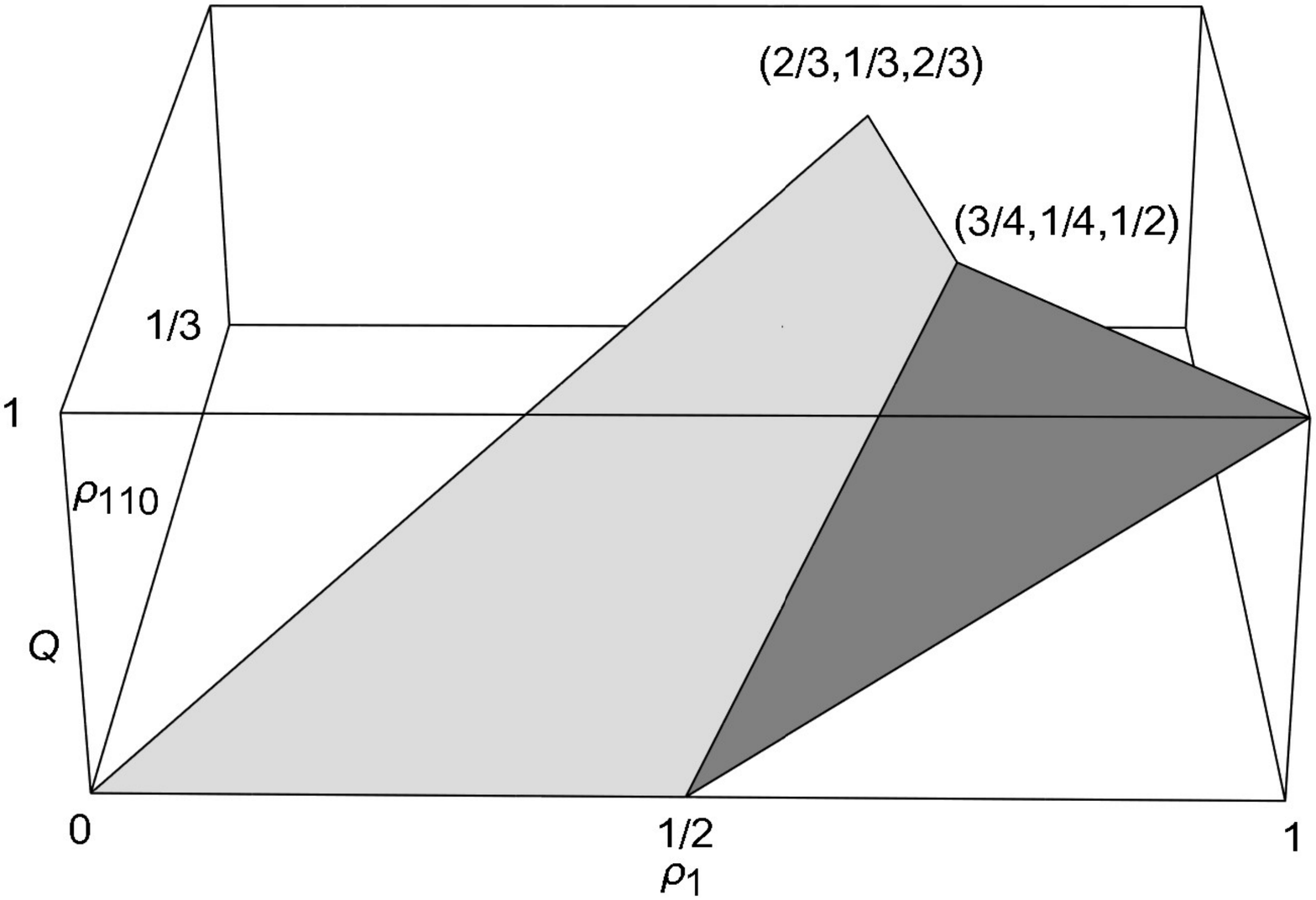}
\caption{Fundamental diagram of (\ref{dsystem}).}
\label{fdd}
\end{center}
\end{figure}

First, we introduce a deterministic discrete particle system described by
\begin{equation}
\label{dsystem}
u_j^{n+1}=u_j^n+q\left(u_{j-2}^n,u_{j-1}^n,u_j^n,u_{j+1}^n\right)-q\left(u_{j-1}^n,u_j^n,u_{j+1}^n,u_{j+2}^n\right).
\end{equation}
The variable $u$ is binary which takes value $0$ or $1$, $j$ is an integer site number and $n$ is an integer time. The function $q$ is a flux written by Table~\ref{flux}.
\begin{table}[h]
\caption{Rule table of $q(w,x,y,z)$ of (\ref{dsystem}). Upper and lower rows denote $(w,x,y,z)$ and $q(w,x,y,z)$ respectively.}
\label{flux}
\begin{center}
$
\begin{array}{r}
\begin{array}{|c||c|c|c|c|c|c|c|c|}
\hline
a,b,c,d
& 1111 & 1110 & 1101 & 1100 & 1011 & 1010 & 1001 & 1000 \\
\hline
q(w,x,y,z)
& 1 & 1 & 1 & 1 & 0 & 0 & 0 & 0 \\
\hline
\end{array}
\medskip\\
\begin{array}{|c|c|c|c|c|c|c|c|}
\hline
0111 & 0110 & 0101 & 0100 & 0011 & 0010 & 0001 & 0000 \\
\hline
0 & 1 & 0 & 0 & 0 & 0 & 0 & 0 \\
\hline
\end{array}
\end{array}
$
\end{center}
\end{table}
We assume the periodic boundary condition for space sites with a period $L$. 
From the above evolution equation, it is easily shown that 
\begin{equation}
\label{law}
\sum_{j=1}^{L}u_j^{n+1}=\sum_{j=1}^{L}u_j^n, \ \ \ \sum_{j=1}^{L}{u_{j-1}^{n+1}u_{j}^{n+1}\left(1-u_{j+1}^{n+1}\right)}=\sum_{j=1}^{L}{u_{j-1}^{n}u_{j}^{n}\left(1-u_{j+1}^{n}\right)}.
\end{equation}
We use a notation $m_{x_1 x_2\cdots x_k}$ which expresses the number of local patterns of $x_1 x_2\cdots x_k$ over $L$ sites. The notation $\rho$ is a density of those patterns ($\rho_{x_1 x_2\cdots x_k}=m_{x_1 x_2\cdots x_k}/L$). Thus, above laws in (\ref{law}) show that $\rho_1$ and $\rho_{110}$ are conserved quantities.
Mean flux over the space sites of the asymptotic behavior is defined by
\begin{equation}
Q=\lim_{n\to \infty}{\frac{1}{L}\sum_{j=1}^{L}q\left(u_{j-2}^n,u_{j-1}^n,u_j^n,u_{j+1}^n\right)}.
\end{equation}
Table~\ref{flux} means the following motion rule of particles.
\begin{itemize}
\item
An isolated particle (010) does not move.
\item
For a pair of adjacent two particles (0110), both particles move.
\item
For a sequence of more than two particles ($011\ldots10$), particles other than the leftmost move.
\end{itemize}
Figure~\ref{te1} shows an example of solution to this system. 
The mean flux Q, which is a mean momentum of particles, depends uniquely on a pair of the densities $\rho_1$ and $\rho_{110}$ as follows\cite{3d}:
\begin{equation}
Q=\max{\left(2\rho_1-1, 2\rho_{110} \right)}.
\label{qd}
\end{equation}
Figure~\ref{fdd} shows the three-dimensional `fundamental diagram' obtained by (\ref{qd}). The domain $(\rho_1, \rho_{110})$ is restricted by $2\rho_{110} \le \rho_1 \le 1-\rho_{110}$ considering the relation between $m_1$ and $m_{110}$. The usual fundamental diagram is defined by the relation between mean flux and density. Since mean flux of this system depends on two independent quantities ($\rho_1$ and $\rho_{110}$), the diagram becomes three-dimensional.
\par
Second, we introduce a stochastic parameter to the above deterministic system and propose a conjecture about asymptotic distribution of the stochastic system. Based on the conjecture, expected values of the mean flux can be derived. Moreover, we confirm that the theoretical formula (\ref{qd}) in the deterministic case can be derived by the limit of the stochastic parameter.\par
Contents of this article are as follows: In the section \ref{2}, we propose stochastic extension of the particle system (\ref{flux}) and derive an explicit formula of mean flux of the asymptotic behavior. In the section \ref{3} we introduce a conjecture about asymptotic distribution of the stochastic system which is derived by eigenvalue problems of transition matrices\cite{endo}. Finally, we take the limit of stochastic parameter to obtain the deterministic profile of the fundamental diagram shown in Figure~\ref{flux}.

\section{Stochastic particle system} \label{2}
We introduce an external variable $a$ into Table~\ref{flux} preserving the conservation laws of (\ref{law}) for $\rho_1$ and $\rho_{110}$, and obtain a stochastic particle system given by equation (\ref{eqs}) and Table~\ref{fluxs}. 
\begin{equation}
\label{eqs}
u_j^{n+1}=u_j^n+q\left(u_{j-2}^n,u_{j-1}^n,u_j^n,u_{j+1}^n\right)-q\left(u_{j-1}^n,u_j^n,u_{j+1}^n,u_{j+2}^n\right).
\end{equation}
\begin{table}[h]
\caption{Rule table of $q(w,x,y,z)$ of (\ref{eqs}).}
\label{fluxs}
\begin{center}
$
\begin{array}{r}
\begin{array}{|c||c|c|c|c|c|c|c|c|}
\hline
(w,x,y,z)
& 1111 & 1110 & 1101 & 1100 & 1011 & 1010 & 1001 & 1000 \\
\hline
q(w,x,y,z)
& 1 & 1 & 1 & 1 & 0 & 0 & 0 & 0 \\
\hline
\end{array}
\medskip\\
\begin{array}{|c|c|c|c|c|c|c|c|}
\hline
0111 & 0110 & 0101 & 0100 & 0011 & 0010 & 0001 & 0000 \\
\hline
a & 1 & 0 & 0 & 0 & 0 & 0 & 0 \\
\hline
\end{array}
\end{array}
$
\end{center}
\end{table}
The variable $a$ is a stochastic parameter defined by
\begin{equation}
a=
\begin{cases}
0 & \text{(with probability $\alpha$)} \\
1 & \text{(with probability $1-\alpha$).}
\end{cases}
\end{equation}

\begin{figure}[h]
\begin{center}
\setlength\unitlength{1truecm}
\begin{picture}(4,4.5)(1.5,0)
\put(-0.27,4.3){\vector(1,0){2}}
\put(2,4.2){$j$}
\put(-0.27,4.3){\vector(0,-1){2}}
\put(-0.35,1.93){$n$}
\put(0,0){\includegraphics[width=70mm]{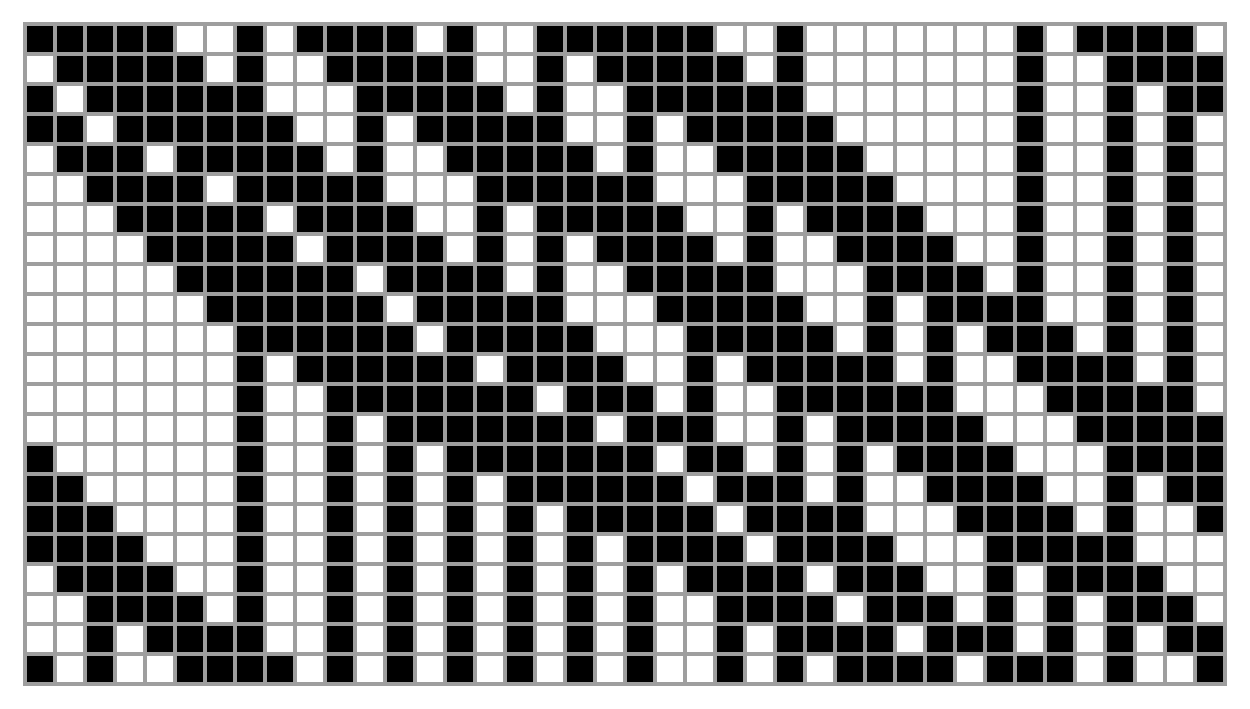}}
\end{picture}
\caption{Example of time evolution of (\ref{eqs}) for $\alpha=0.5$.}
\label{ste}
\end{center}
\end{figure}
\begin{figure}[h]
\begin{center}
\includegraphics[width=80mm]{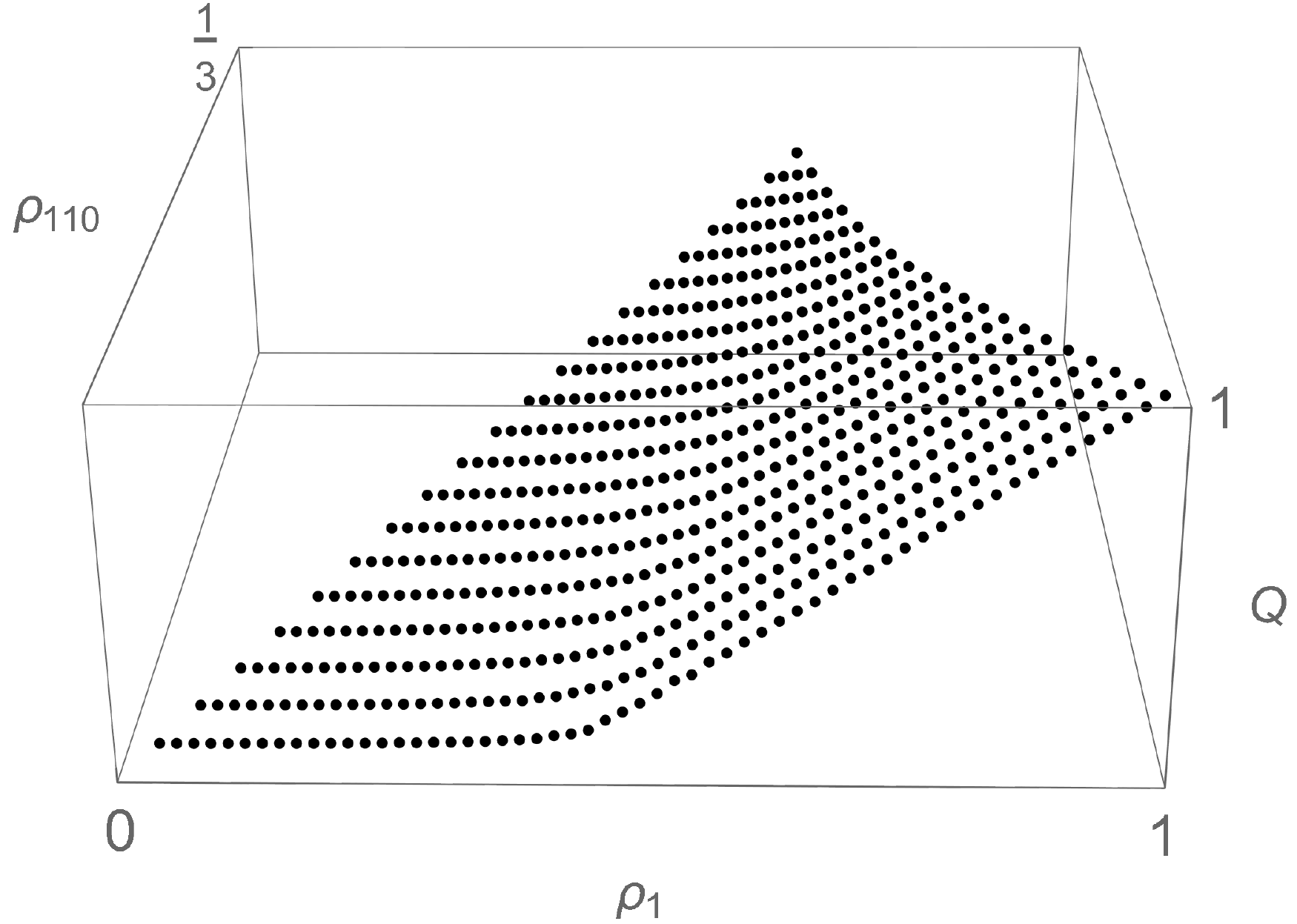}
\caption{Numerical results of fundamental diagram of (\ref{eqs}) for $L=60, \ \alpha=0.7$ averaged from $n=0$ to 3000.}
\label{sim}
\end{center}
\end{figure}
\begin{figure}[h]
\begin{center}
\includegraphics[width=120mm]{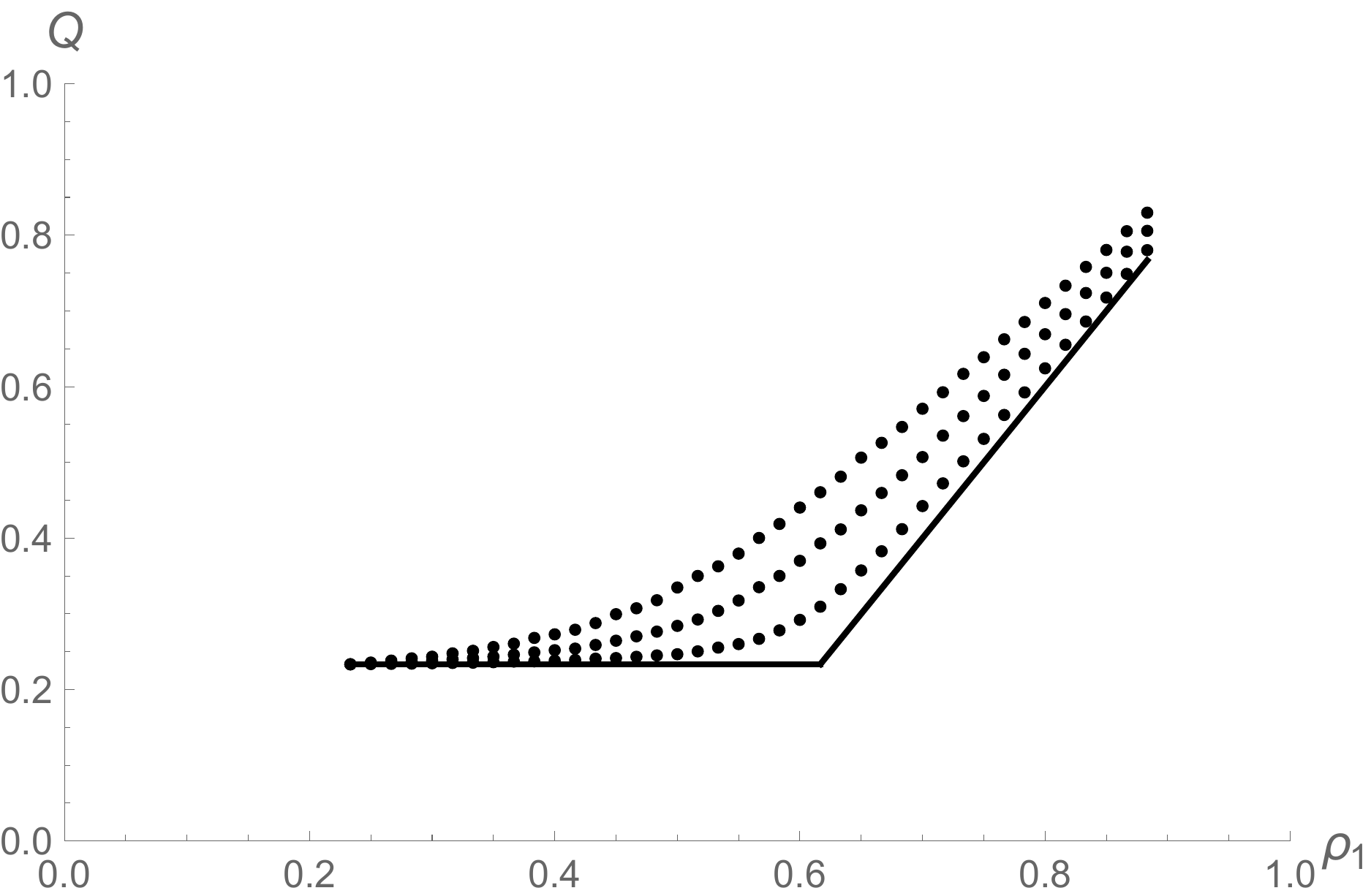}
\caption{Example of fundamental diagram of (\ref{eqs}) for $L=60, \ \rho_{110}=7/60, \ 7/30 \le \rho_1 \le 53/60$. Dots are numerical results for $\alpha=0.5, \ 0.7, \ 0.9$ averaged from $n=0$ to 3000, and the curve is for $\alpha=1$, which is obtained by (\ref{qd}).}
\label{ss}
\end{center}
\end{figure}
Figure~\ref{ste} shows an example of time evolution, and Figure~\ref{sim} shows numerical results of mean flux of the stochastic system.
Comparing with Figure~\ref{fdd}, surface of mean flux of the stochastic system converges to that of the deterministic system (\ref{flux}) along $\alpha \to 1$. Figure~\ref{ss} shows a sectional graph of $\rho_{110}=7/60$ and $\alpha=0.5, \ 0.7, \ 0.9, \ 1$. 
\par
In order to analyze this stochastic system, we express the time evolution using a new variable $v_j^n$ given by $v_j^n=u_{n-j}^n$. The equation is
\begin{equation}
\label{eqsg}
v_j^{n+1}=v_j^n+q\left(v_{j-3}^n,v_{j-2}^n,v_{j-1}^n,v_{j}^n\right)-q\left(v_{j-2}^n,v_{j-1}^n,v_{j}^n,v_{j+1}^n\right),
\end{equation}
and $q(w,x,y,z)$ is given by Table~\ref{fluxsg}
\begin{table}[h]
\caption{Rule table of $q(w,x,y,z)$ of (\ref{eqsg})}
\label{fluxsg}
\begin{center}
$
\begin{array}{r}
\begin{array}{|c||c|c|c|c|c|c|c|c|}
\hline
(w,x,y,z)
& 1111 & 1110 & 1101 & 1100 & 1011 & 1010 & 1001 & 1000 \\
\hline
q(w,x,y,z)
& 0 & b & 0 & 0 & 0 & 1 & 0 & 0 \\
\hline
\end{array}
\medskip\\
\begin{array}{|c|c|c|c|c|c|c|c|}
\hline
0111 & 0110 & 0101 & 0100 & 0011 & 0010 & 0001 & 0000 \\
\hline
0 & 0 & 0 & 0 & 0 & 1 & 0 & 0 \\
\hline
\end{array}
\end{array}
$
\end{center}
\end{table}
where the variable $b$ is a stochastic parameter defined by
\begin{equation}
b=
\begin{cases}
1 & \text{(with probability $\alpha$)} \\
0 & \text{(with probability $1-\alpha$).}
\end{cases}
\end{equation}
The motion rule of this stochastic system is as follows.
\begin{itemize}
\item
An isolated particle (010) moves to its neighboring right empty site.
\item
For a pair of adjacent two particles (0110), both stay at their positions.
\item
For a sequence of more than two particles ($011\ldots10$), the rightmost particle moves to its neighboring right empty site with probability $\alpha$. Particles other than the rightmost stay at their positions.
\end{itemize}
\begin{figure}[h]
\begin{center}
\setlength\unitlength{1truecm}
\begin{picture}(4,4.5)(1.5,0)
\put(-0.27,4.3){\vector(1,0){2}}
\put(2,4.2){$j$}
\put(-0.27,4.3){\vector(0,-1){2}}
\put(-0.35,1.93){$n$}
\put(0,0){\includegraphics[width=70mm]{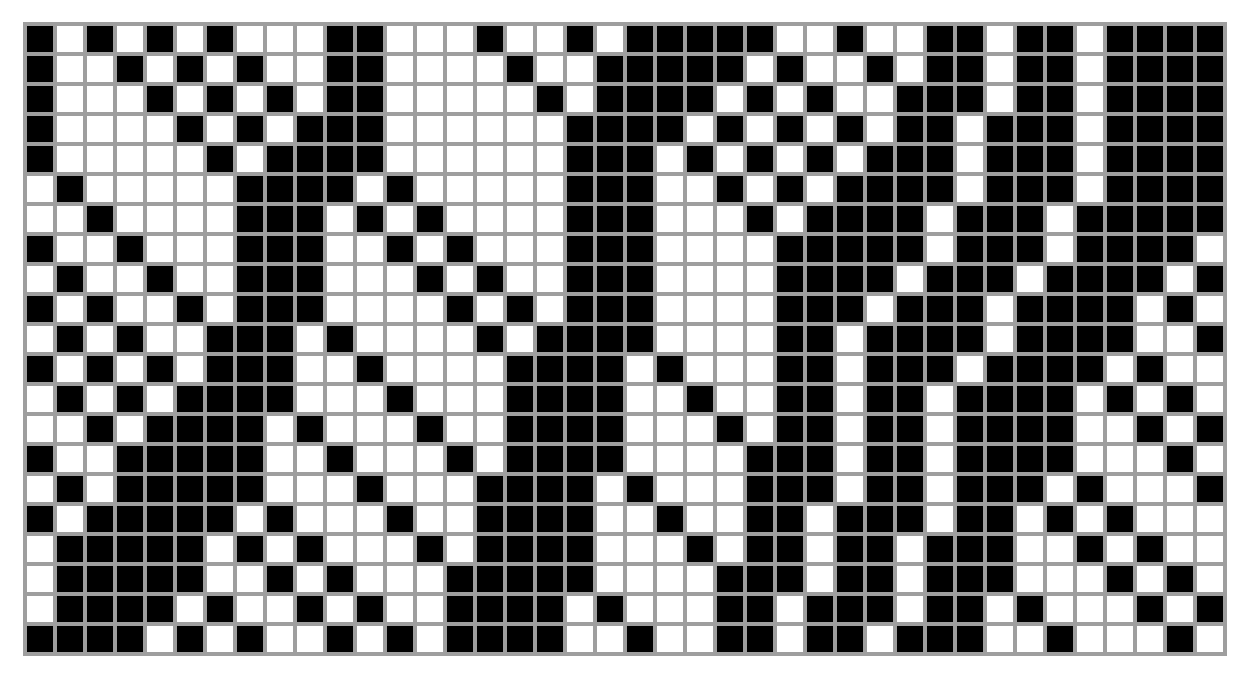}}
\end{picture}
\caption{Example of time evolution of (\ref{eqsg}) for $\alpha=0.5$.}
\label{steg}
\end{center}
\end{figure}
Figure~\ref{steg} shows an example of time evolution of the transformed stochastic system. From the above motion rule of particles, the mean flux of the system can be expressed by the local densities.
\begin{equation}
Q=\lim_{n\to \infty}{\left(\alpha\rho_{1110}^n+\rho_{010}^n\right)}.
\label{Q}
\end{equation}
\section{Asymptotic distribution} \label{3}
Let us consider the stochastic process provided by Table~\ref{fluxsg}. Define a `configuration' by a set of values over all sites at any time. Then it changes to another along the time evolution by Table~\ref{fluxsg}. This transition of configurations can be treated as a stochastic process using the transition matrix describing the probability of their transition. This transition process can be separated by sets of configurations since all of them cannot be obtained by the evolution of a given configuration. Thus, we define an irreducible set $\Omega$ of configurations where any of $\Omega$ can be transferred to all other ones. Moreover, to make the expression of $\Omega$ compact, we identify configurations up to cyclic rotation. \par
\begin{figure}[h]
\begin{center}
\includegraphics[width=120mm]{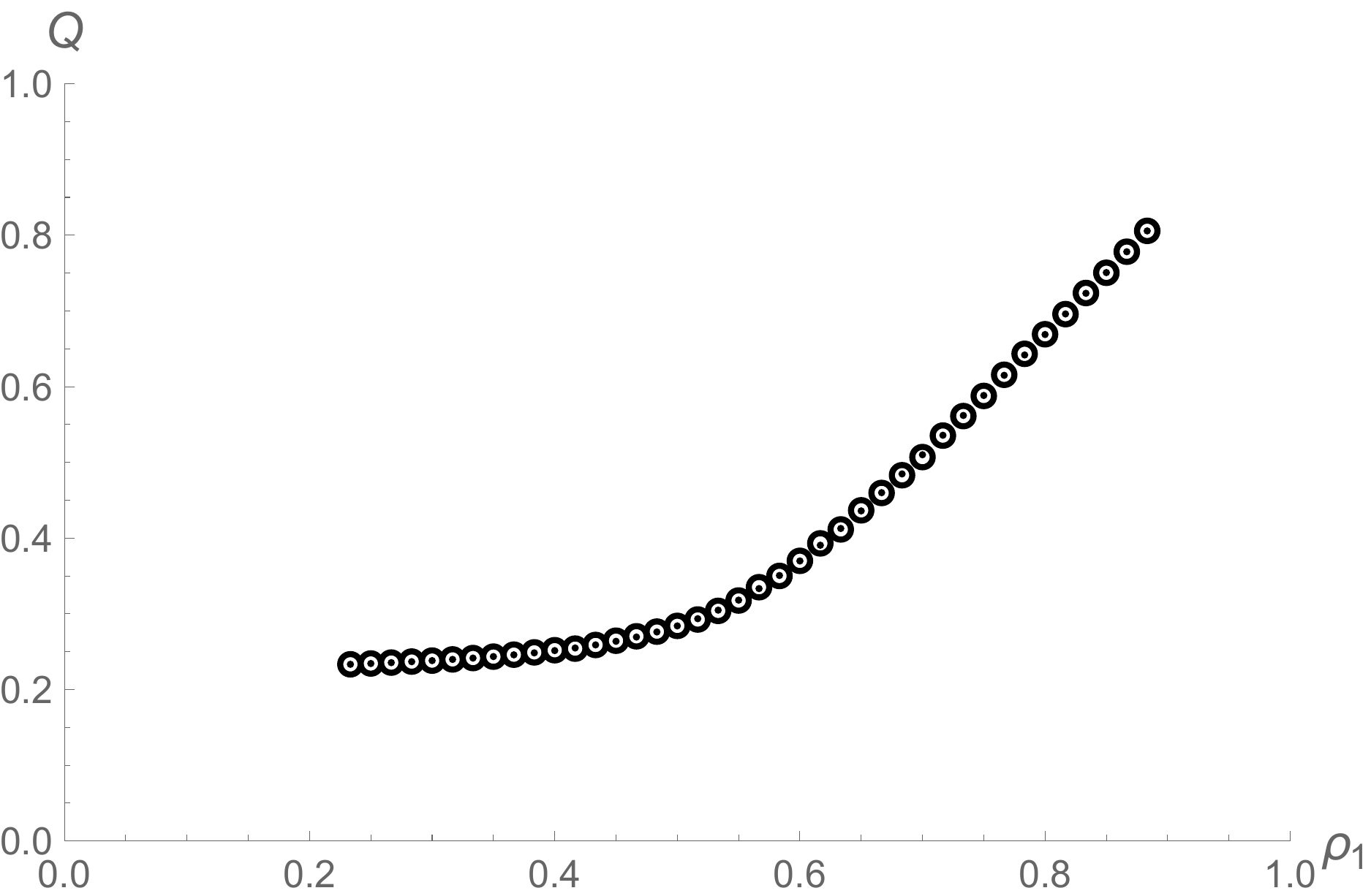}
\caption{Example of fundamental diagram of (\ref{eqs}) for $\rho_{110}=7/60$. Small dots ($\bullet$) are obtained by (\ref{Qs}) for $L=60, \alpha=0.7$. Circles ($\bigcirc$) are numerical results for the same parameters averaged from $n=0$ to 3000.}
\label{fdtc}
\end{center}
\end{figure}

Note that $\Omega$ cannot be determined only by length $L$, two conserved quantities $m_1$ and $m_{110}$. For the same set of $L$, $m_1$ and $m_{110}$, there exist configurations which cannot be transferred each other. For example, for $L=10, m_1=6, m_{110}=2$ there are two irreducible sets $\Omega_1$ and $\Omega_2$ given by
\begin{equation*}
\Omega_1=
\begin{array}{l}
\{0001101111,\ 0001110111,\ 0001111011,\ 0010110111,\\
\ 0010111011,\ 0011011101,\ 0011101101,\ 0101011011\},
\end{array}
\end{equation*}
\begin{equation*}
\Omega_2=
\begin{array}{l}
\{0011001111,\ 0011010111,\ 0011100111,\ 0011101011,\ 0101101011\}.
\end{array}
\end{equation*}
The transition matrix of $\Omega_1$ is
\begin{equation*}
\begin{array}{c}
0001101111\\
0001110111\\
0001111011\\
0010110111\\
0010111011\\
0011011101\\
0011101101\\
0101011011
\end{array}
\left(
\begin{array}{cccccccc}
1-\alpha & 0 & 0 & 0 & 0 & \alpha & 0 & 0 \\
(1-\alpha)\alpha & (1-\alpha)^2 & 0 & 0 & 0 & \alpha^2 & (1-\alpha)\alpha & 0 \\
0 & \alpha & 1-\alpha & 0 & 0 & 0 & 0 & 0 \\
0 & 1-\alpha & 0 & 0 & 0 & 0 & \alpha & 0 \\
0 & \alpha & 1-\alpha & 0 & 0 & 0 & 0 & 0 \\
0 & 0 & 0 & 1-\alpha & 0 & 0 & 0 & \alpha \\
0 & 0 & 0 & \alpha & 1-\alpha & 0 & 0 & 0 \\
0 & 0 & 0 & 0 & 1 & 0 & 0 & 0 \\
\end{array}
\right),
\end{equation*}
where each component of the transition matrix ($a_{i,j}$) is a transition probability from the $i$ th configuration to the $j$ th configuration which is determined by the local dynamics.
An eigenvector of the matrix for eigenvalue 1 is
\begin{equation*}
(\frac{1-\alpha}{\alpha^2},\ \frac{1}{\alpha^2},\frac{1-\alpha}{\alpha^2},\ \frac{1}{\alpha},\ \frac{1}{\alpha},\ \frac{1}{\alpha},\ \frac{1}{\alpha},\ 1).
\end{equation*}
Supposing that the time evolution is ergodic, the above eigenvector shows ratio of probability of which each configuration occurs in the asymptotic state. The transition matrix of $\Omega_2$ is
\begin{equation*}
\begin{array}{c}
0011001111\\
0011010111\\
0011100111\\
0011101011\\
0101101011\\
\end{array}
\left(
\begin{array}{ccccc}
1-\alpha & 0 & 0 & \alpha & 0 \\
1-\alpha & 0 & 0 & \alpha & 0 \\
0 & 2\alpha(1-\alpha) & (1-\alpha)^2 & 0 & \alpha^2 \\
0 & \alpha & 1-\alpha & 0 & 0 \\
0 & 0 & 1 & 0 & 0 \\
\end{array}
\right),
\end{equation*}
and its eigenvector of eigenvalue 1 is 
\begin{equation*}
(\frac{2(1-\alpha)}{\alpha^2},\ \frac{2}{\alpha},\frac{1}{\alpha^2},\ \frac{2}{\alpha},\ 1).
\end{equation*}
We derive an example of eigenvector for another case of $L=16, m_1=11, m_{110}=4$.\\
One of sets of configurations is
\begin{equation*}
\begin{array}{l}
\{0011011011011111,\ 0011011011101111,\ 0011011011110111,\\
\ 0011011011111011,\ 0011011101101111,\ 0011011101110111,\\
\ 0011011101111011,\ 0011011110110111,\ 0011011110111011,\\
\ 0011011111011011,\ 0011101101101111,\ 0011101101110111,\\
\ 0011101101111011,\ 0011101110110111,\ 0011101110111011,\\
\ 0011101111011011,\ 0011110110110111,\ 0011110110111011,\\
\ 0011110111011011,\ 0011111011011011,\ 0101101101101111,\\
\ 0101101101110111,\ 0101101101111011,\ 0101101110110111,\\
\ 0101101110111011,\ 0101101111011011,\ 0101110110110111,\\
\ 0101110110111011,\ 0101110111011011,\ 0101111011011011\}.
\end{array}
\end{equation*}
The eigenvector for eigenvalue 1 of the transition matrix is 
\begin{eqnarray*}
(\frac{1-\alpha}{\alpha},\ \frac{1}{\alpha},\ \frac{1}{\alpha},\ \frac{1-\alpha}{\alpha},\ \frac{1}{\alpha},\ \frac{1}{(1-\alpha)\alpha},\ \frac{1}{\alpha},\ \frac{1}{\alpha},\ \frac{1}{\alpha},\ \frac{1-\alpha}{\alpha},\ \frac{1}{\alpha},\ \frac{1}{(1-\alpha)\alpha},\ \frac{1}{\alpha},\ \frac{1}{(1-\alpha)\alpha},\ \\
\frac{1}{(1-\alpha)\alpha},\ \frac{1}{\alpha},\ \frac{1}{\alpha},\ \frac{1}{\alpha},\ \frac{1}{\alpha},\ \frac{1-\alpha}{\alpha},\ 1,\ \frac{1}{1-\alpha},\ 1,\ \frac{1}{1-\alpha},\ \frac{1}{1-\alpha},\ 1,\ \frac{1}{1-\alpha},\ \frac{1}{1-\alpha},\ \frac{1}{1-\alpha},\ 1).
\end{eqnarray*}
Examining other concrete examples, we propose a conjecture for the asymptotic distribution of configurations as follows. 
\begin{quote}
Conjecture: For arbitrary $\Omega$, the probability of configuration $x \in\Omega$ in the asymptotic state is given by
\begin{equation}
p\left(x\right)\propto \frac{\alpha^{m_{010}\left(x\right)}}{\left(1-\alpha\right)^{m_{1110}\left(x\right)+m_{010}\left(x\right)}},
\label{ratio}
\end{equation}
where $m_{1110}\left(x\right)$ and $m_{010}\left(x\right)$ is the number of local patterns 1110 and 010 in the configuration $x$.
\end{quote}
By the above conjecture, the probability of $x$ in the asymptotic state is
\begin{equation}
p\left(x\right)=\frac{\frac{\alpha^{m_{010}\left(x\right)}}{\left(1-\alpha\right)^{m_{1110}\left(x\right)+m_{010}\left(x\right)}}}{\sum_{k_1,k_2}{\frac{\alpha^{k_2}}{\left(1-\alpha\right)^{k_1+k_2}}N\left(k_1,k_2\right)}},
\label{dist}
\end{equation}
where $N(k_1,k_2)$ is a partition function, that is, the number of configurations $x$ satisfying $m_{1110}(x)=k_1$ and $m_{010}(x)=k_2$. We obtain that the maximum value of $k_1+k_2$, that is, the maximum value of the sum of numbers of local patterns 1110 and 010 in $\Omega$, is $\min{\left(L-m_1,m_1-2m_{110}\right)}$\cite{3d}. Since the mean flux $Q$ is an expected value of $\lim_{n\to \infty}{\left(\alpha\rho_{1110}^n+\rho_{010}^n\right)}$, we have
\begin{equation}
\label{Qs}
Q=\frac{1}{L}\frac{\sum_{0\le k_1+k_2\le\min{\left(L-m_1,m_1-2m_{110}\right)}}{\left(\alpha k_1+k_2\right)\frac{\alpha^{k_2}}{\left(1-\alpha\right)^{k_1+k_2}}}N\left(k_1,k_2\right)}{\sum_{0\le k_1+k_2\le\min{\left(L-m_1,m_1-2m_{110}\right)}}{\frac{\alpha^{k_2}}{\left(1-\alpha\right)^{k_1+k_2}}}N\left(k_1,k_2\right)}.
\end{equation}
Using inverse transformation $u_{n-j}^n=v_j^n$, the mean flux of (\ref{eqs}) is $\rho_1-Q$. Figure~\ref{fdtc} shows a comparison between theoretical values by (\ref{Qs}) and numerical values.
\par
Since the terms $\alpha^{k_2}/\left(1-\alpha\right)^{k_1+k_2}$ of maximum $k_1+k_2$ remains in the limit $\alpha \to 1$ for (\ref{Qs}), we have
\begin{eqnarray*}
\lim_{\alpha \to 1} (\rho_1-Q)&=&\frac{m_1}{L}-\frac{1}{L}\frac{\min{\left(L-m_1,m_1-2m_{110}\right)}\sum_{k_1+k_2=\min{\left(L-m_1,m_1-2m_{110}\right)}}{N\left(k_1,k_2\right)}}{\sum_{k_1+k_2=\min{\left(L-m_1,m_1-2m_{110}\right)}}{N\left(k_1,k_2\right)}} \\
&=&\frac{m_1-\min{\left(L-m_1,m_1-2m_{110}\right)}}{L}=\rho_1-\min{\left(1-\rho_1,\rho_1-2\rho_{110}\right)}\\
&=&\rho_1+\max{\left(\rho_1-1, 2\rho_{110}-\rho_1\right)}=\max{\left(2\rho_1-1, 2\rho_{110} \right)}.
\end{eqnarray*}
It coincides with (\ref{qd}) of the deterministic system (\ref{dsystem}).
\section{Conclusion} \label{4}
We analyzed asymptotic behavior of the stochastic particle system of 5 neighbors given by (\ref{eqs}) and (\ref{eqsg}). Considering the irreducible sets of configurations, we proposed a conjecture about components of eigenvector of transition matrix for eigenvalue 1 is determined only by the number of the local patterns 1110 and 010 depending on each configuration. Using this conjecture, we derived asymptotic distribution of configurations. Then, we derived mean flux of the stochastic system as an expected value of the densities of local patterns 1110 and 010. The three-dimensional fundamental diagram can be obtained theoretically by (\ref{Qs}), and it coincides with the numerical results. Moreover, we derived theoretical formula of mean flux of the deterministic system through the limit $\alpha \to 1$. \par
The partition function $N(k_1,k_2)$ is the number of configurations of which the number of local patterns 1110 and 010 are $k_1$ and $k_2$. Since it is hard to enumerate configurations with specific local patterns for arbitrary set $\Omega$, general formula of the partition function has not been obtained yet. However, once $\Omega$ is given concretely, $N(k_1,k_2)$ can be easily calculated, thus we confirmed the theoretical formula of the fundamental diagram. Derivation of the general formula of the partition function is one of the future problems. Moreover, another future problem of our theory is to prove the configuration of the asymptotic distribution (\ref{ratio}). \par
Partition function and fundamental diagram are generally hard to derive for arbitrary stochastic particle systems. Our results reported in this article imply a breakthrough to solve this difficulty. The key idea is to introduce two or more quantities for classification of asymptotic set of configurations and to derive the distribution using them. Since this idea can be applied to other systems, it is interesting to obtain a general formulation of stochastic particle systems with a common mechanism.


\begin{thebibliography}{15}
%
\bibitem{fuks}
H.~Fuk\'s,
Critical behaviour of number-conserving cellular automata with nonlinear fundamental diagrams, J. Stat. Mech. (2004), P07005.
%
\bibitem{nishinari}
K.~Nishinari and D.~Takahashi,
Analytical properties of ultradiscrete Burgers equation and rule-184 cellular automaton,
J. Phys. A. {\bf 31} (1998), 5439--5450.
%
\bibitem{spitzer}
F.~Spitzer, Interaction of Markov Processes, Adv. Math. {\bf 5} (1970), 246--190.
%
\bibitem{derrida1}
B.~Derrida, E.~Domany and E.~Mukamel, An exact solution of a one-dimensional asymmetric exclusion model with open boundaries, J. Stat. Phys. {\bf 69} (1992), 667--687. 
%
\bibitem{derrida2}
B.~Derrida, M. R.~Evans, V.~Hakim and V.~Pasquier, Exact solution of a 1D asymmetric exclusion model using a matrix formulation, J. Phys. A: Math. Gen. {\bf 26} (1993), 1493--1517. 
%
\bibitem{sasamoto}
T.~Sasamoto, One-dimensional partially asymmetric simple exclusion process with open boundaries: Orthogonal polynomials approach, J. Phys. A: Math. Gen. {\bf32} (1999), 7109--7131. 
%
\bibitem{nagel}
K.~Nagel and M.~Schreckenberg, A cellular automaton model for freeway traffic, J. Phys. I France. {\bf 2} (1992), 2221--2229.
%
\bibitem{ns}
M.~Schreckenberg, A.~Schadschneider, K.~Nagel and N.~Ito, Discrete stochastic models for traffic flow, Phys. Rev. E. {\bf 51} (1995), 2939--2949. 
%
\bibitem{3d}
K.~Endo and D.~Takahashi, Three-dimensional fundamental diagram of particle system of 5 neighbors with two conserved densities, arXiv preprint: 2112.12929.
%
\bibitem{endo}
K.~Endo, New approach to evaluate the asymptotic distribution of particle systems expressed by probabilistic cellular automata, Japan J. Indust. Appl. Math. {\bf 37} (2020), 461--484.
\end{thebibliography}
\end{document}